\documentclass[aps, prl, twocolumn,superscriptaddress,nofootinbib,10pt]{revtex4-1}

\usepackage{graphicx}
\usepackage{dcolumn}
\usepackage{amssymb,amsmath,amsthm,mathrsfs}
\usepackage{epstopdf}
\usepackage{lipsum}
\usepackage{hyperref}
\usepackage{empheq}
\usepackage{color}
\usepackage[normalem]{ulem} 
\usepackage{physics}
\usepackage[normalem]{ulem}
\usepackage[ruled,vlined]{algorithm2e}

\newcommand{\be}{\begin{equation}}
\newcommand{\ee}{\end{equation}}
 \newcommand{\bea}{\begin{eqnarray}}
\newcommand{\eea}{\end{eqnarray}}

\newcommand{\el}{\nonumber \\}

\newcommand{\pat}{\partial}

\renewcommand{\a}{\alpha}
\renewcommand{\b}{\beta}

\renewcommand{\d}{\delta}

\newcommand{\GN}{G_{\mathrm{N}}}

\newcommand{\rmd}{\mathrm{d}}

\newcommand{\thr}{\mathrm{th}}

\newcommand{\ie}{i.e.\ }
\newcommand{\eg}{e.g.\ }

\newcommand{\av}[1]{\langle{#1}\rangle}

\newcommand{\mr}{\mathrm}

\newcommand{\Pzeta}{\mathcal{P}_\zeta}
\newcommand{\Pphi}{\mathcal{P}_{\d\phi}}
\newcommand{\kpbh}{k_\mr{PBH}}
\newcommand{\bk}{\mathbf{k}}
\newcommand{\bp}{\mathbf{p}}
\newcommand{\bx}{\mathbf{x}}
\newcommand{\kc}{k_\sigma}
\newcommand{\hxi}{{\hat{\xi}}}
\newcommand{\tk}{{\tilde{k}}}
\newcommand{\kini}{k_\text{ini}}
\newcommand{\kend}{k_\text{end}}

\newcommand{\fig}[1]{figure \ref{#1}}

\newcommand{\C}{\mathcal{C}}
\newcommand{\avC}{\bar{\mathcal{C}}}

\begin{document}

\rightline{HIP-2023-18/TH}
\vspace*{0.5cm}

\title{Primordial black hole compaction function from \\ stochastic fluctuations in ultra-slow-roll inflation}

\newcommand{\addressHIP}{University of Helsinki, Helsinki Institute of Physics, P.O. Box 64, FIN-00014 University of Helsinki, Finland}
\newcommand{\addressHU}{University of Helsinki, Helsinki Institute of Physics and Department of Physics, P.O. Box 64, FIN-00014 University of Helsinki, Finland}
\newcommand{\addressTallinn}{Laboratory of High Energy and Computational Physics, National Institute of Chemical Physics and Biophysics, R\"{a}vala puiestee 10, 10143 Tallinn, Estonia}
\newcommand{\addressLancaster}{Consortium for Fundamental Physics, Physics Department, Lancaster University,
Lancaster LA1 4YB, United Kingdom}

\author{Sami Raatikainen} \affiliation{\addressHIP}
\author{Syksy R\"{a}s\"{a}nen} \affiliation{\addressHU}
\author{Eemeli Tomberg} \affiliation{\addressTallinn} \affiliation{\addressLancaster}

\date{\today}

\begin{abstract}

We study the formation of primordial black holes (PBH) with ultra-slow-roll inflation when stochastic effects are important. We use the $\Delta N$ formalism and simplify the stochastic equations with an analytical constant-roll approximation. Considering a viable inflation model, we find the spatial profile of the PBH compaction function numerically for each stochastic patch, without assumptions about Gaussianity or the radial profile. The stochastic effects that lead to an exponential tail for the density distribution also make the compaction function very spiky, unlike assumed in the literature. Naively using collapse thresholds found for smooth profiles, the PBH abundance is enhanced by up to a factor of $10^9$, and the PBH mass distribution is spread over three orders of magnitude in mass. The results point to a need to redo numerical simulations of PBH formation with spiky profiles.

\end{abstract}


\maketitle

\raggedbottom 

{\it Introduction.--}
Primordial black holes (PBHs) \cite{Hawking:1971ei, Chapline:1975, Dolgov:1992pu, Ivanov:1994pa, Yokoyama:1995ex, GarciaBellido:1996qt, Jedamzik:1996mr, Ivanov:1997ia, Blais:2002nd} can constitute all of the dark matter or seed supermassive black holes \cite{Carr:2020gox, Carr:2020xqk, Green:2020jor}. Generating PBHs requires large density fluctuations. They may originate in the same process as the small fluctuations seen in the cosmic microwave background (CMB). The most successful proposal for this is cosmic inflation, in the simplest case driven by one scalar field, the inflaton \cite{Starobinsky:1979ty, Starobinsky:1980te, Kazanas:1980tx, Guth:1980zm, Sato:1980yn, Mukhanov:1981xt, Linde:1981mu, Albrecht:1982wi, Hawking:1981fz, Chibisov:1982nx, Hawking:1982cz, Guth:1982ec, Starobinsky:1982ee, Sasaki:1986hm, Mukhanov:1988jd}. The curvature perturbation $\zeta$ is inversely proportional to the inflaton velocity, so a feature in the inflaton potential that slows down the field amplifies $\zeta$. A much studied case is ultra-slow-roll (USR) inflation, where the classical force due to the potential is small or opposes the inflaton velocity, and the velocity decreases exponentially \cite{Ivanov:1994pa, Faraoni:2000vg, Kinney:2005vj, Martin:2012pe, Garcia-Bellido:2017mdw, Ezquiaga:2017fvi, Kannike:2017bxn, Germani:2017bcs, Motohashi:2017kbs, Dimopoulos:2017ged, Gong:2017qlj, Ballesteros:2017fsr, Hertzberg:2017dkh, Pattison:2018, Biagetti:2018pjj, Ezquiaga:2018gbw, Rasanen:2018fom}.

A simple way to determine the PBH abundance is to solve for the inflaton background evolution, calculate the linear perturbations around it to obtain the curvature power spectrum ${\mathcal P}_{\zeta}$, and, assuming Gaussian statistics, use ${\mathcal P}_{\zeta}$ to find the fraction of values of $\zeta$ that exceed the curvature threshold $\zeta_\thr$ corresponding to PBH formation \cite{Garcia-Bellido:2017mdw, Ezquiaga:2017fvi, Kannike:2017bxn, Germani:2017bcs, Motohashi:2017kbs, Rasanen:2018fom}. This approach has two shortcomings.

First, stochastic effects due to the stretching of short-wavelength modes into long wavelengths couple the evolution of the background to the perturbations \cite{Starobinsky:1986fx, Morikawa:1989xz, Habib:1992ci, Mijic:1994vv, Starobinsky:1994bd}. USR is particularly sensitive to stochastic noise, because the classical field velocity is small and perturbations are large \cite{Pattison:2017mbe, Biagetti:2018pjj, Ezquiaga:2018gbw, Cruces:2018cvq, Firouzjahi:2018vet, Pattison:2019hef, Ezquiaga:2019ftu, Ballesteros:2020sre, De:2020hdo, Firouzjahi:2018vet, Figueroa:2020jkf, Rigopoulos:2021nhv, Pattison:2021oen, Figueroa:2021zah, Tomberg:2022mkt, Tomberg:2023kli, Mishra:2023lhe}. Stochastic kicks lead to an exponential tail for the fluctuations, which can dominate the linear theory Gaussian tail \cite{Pattison:2017mbe, Ezquiaga:2019ftu, Figueroa:2020jkf, Pattison:2021oen, Figueroa:2021zah, Tomberg:2022mkt, Tomberg:2023kli}.\footnote{Classical non-linearities can also lead to an exponential or heavier tail \cite{Biagetti:2021eep,  Kitajima:2021fpq, Hooshangi:2021ubn, Cai:2021zsp, DeLuca:2022rfz, Ferrante:2022mui, Gow:2022jfb, Pi:2022ysn, Hooshangi:2023kss}.}

It is therefore not enough to consider the power spectrum. In \cite{Figueroa:2020jkf} we presented the first numerical calculation of the non-Gaussian tail of the probability distribution of $\zeta$ due to stochastic effects in USR, using the formation threshold $\zeta_\thr=1$. This brings us to the second problem: $\zeta$ is not a good measure of PBH formation. The curvature perturbation is sensitive to long-wavelength fluctuations, whereas gravitational collapse and formation of a trapped surface is determined by local conditions \cite{Nakama:2013ica, Young:2014ana, Yoo:2018kvb}.

The maximum value of the compaction function $\C\equiv2\GN\Delta M/R$ or its average has been found to be a good indicator of PBH formation \cite{Shibata:1999zs, Harada:2013epa, Harada:2015yda, Musco:2018rwt, Young:2019yug, Young:2019osy, Escriva:2019nsa, Escriva:2019phb, Atal:2019erb, Yoo:2020lmg, Escriva:2020tak, Musco:2020jjb, Escriva:2021aeh, Escriva:2022pnz, Yoo:2022mzl, Harada:2023ffo, Escriva:2023uko, Escriva:2023qnq}. Here $\GN$ is Newton's constant, $\Delta M(t,r)$ is mass excess over the background, and $R(t,r)$ is the areal radius at time $t$ and coordinate radius $r$. For a Gaussian field, large overdensities are close to spherically symmetric, and we assume this also in our non-Gaussian case \cite{Bardeen:1985tr}. On scales larger than the Hubble radius, $\C$ is independent of time.

Checking whether $\C(r)$ exceeds the collapse threshold requires knowing the spatial profile of the perturbation. Previous studies have used the mean profile and its variance calculated for Gaussian random fields \cite{Germani:2018jgr, Young:2019osy, Atal:2019cdz, Kalaja:2019uju, Atal:2019erb, Musco:2020jjb, Kitajima:2021fpq, Yoo:2022mzl, Escriva:2023uko}, or ad hoc profiles \cite{Shibata:1999zs, Nakama:2013ica, Harada:2015yda, Musco:2018rwt, Kawasaki:2019mbl, Young:2019yug, Kalaja:2019uju, Escriva:2019nsa, Escriva:2019phb, Yoo:2020lmg, Escriva:2022pnz, Germani:2023ojx, Escriva:2023qnq}. The former approach is unlikely to be accurate when the statistics are highly non-Gaussian, and the ad hoc profiles have been smoothly varying and have had only one or two maxima. When stochastic effects are significant, $\C(r)$ varies rapidly and has many maxima.

In this Letter we for the first time calculate the compaction function $\C(r)$ directly from stochastic inflaton fluctuations, showing how the stochastic effects that generate the exponential tail of the perturbations also make their spatial profile spiky. We use $\C(r)$ to determine PBH abundance as a function of the collapse threshold, providing a complete route from the inflaton potential to PBH abundance including stochastic effects, without assumptions about the radial profile or Gaussianity. 

\flushbottom 

{\it The compaction function.--}
As in \cite{Figueroa:2020jkf, Figueroa:2021zah}, we use the $\Delta N$ formalism, where $\zeta$ is equal to the difference between the number of e-folds $N$ of a local patch of the universe and the mean number of e-folds $\bar{N}$, $\zeta = N - \bar{N} \equiv \Delta N$ \cite{Sasaki:1995aw, Sasaki:1998ug, Wands:2000dp, Lyth:2004gb}. By simulating one patch of comoving scale $\bk$, we obtain the coarse-grained curvature perturbation $\zeta_{<k} \equiv \int \frac{\rmd^3 p}{(2\pi)^{3/2}} \zeta_\bp e^{i\bp\cdot\bx} \theta(k-p)$, where $k\equiv|\bk|$, $p\equiv|\bp|$. Given spherical symmetry, the Fourier modes satisfy
\bea \label{zeta_k}
    \zeta_\bk = \zeta_k = \frac{\sqrt{2\pi}}{2k^3} \frac{\rmd \zeta_{<k}}{\rmd \ln k} \ .
\eea
We end the simulations on super-Hubble scales, where $\zeta_k$ is time-independent. We probe a large number of $k$ values in the same patch to reconstruct the function $\zeta(r)$ with the inverse Fourier transform:
\bea \label{zeta}
  \zeta(r) &=& \frac{1}{(2\pi)^{3/2}} \int \rmd^3 k \, \zeta_\bk e^{i\bk\cdot\bx} \el
  &=& \sqrt{\frac{2}{\pi}} \int_0^{\infty} \rmd k \, k^2 \zeta_k \frac{\sin(k r)}{k r} \ .
\eea
A window function is sometimes introduced in \eqref{zeta} to cut off scales and define the PBH mass \cite{Ando:2018qdb, Young:2019osy, Kalaja:2019uju}. However, its physical motivation is unclear. Using a Gaussian window function would not have a large effect on our results. Our PBH masses are determined by where the stochastic process produces maxima of $\C(r)$, not fixed a priori. The patches just have to be small enough to be approximated as spherically symmetric, and large enough to capture all maxima that can lead to PBH formation. We also do not include the transfer function, which has a small impact on sub-Hubble evolution \cite{Kalaja:2019uju, Musco:2020jjb, Germani:2023ojx, DeLuca:2023tun}.

On super-Hubble scales in the comoving gauge, the compaction function in the gradient approximation is \cite{Harada:2015yda}
\bea \label{C}
  \mathcal C(r) &=& \frac{2}{3} [ 1 - ( 1 + r \zeta' )^2 ] \ ,
\eea
where $'\equiv\frac{\pat}{\pat r}$. The prefactor is gauge-dependent \cite{Shibata:1999zs, Harada:2015yda, Yoo:2022mzl, Harada:2023ffo}. It has been argued that the areal volume average of $\C$ is a better indicator of collapse \cite{Escriva:2019phb, Atal:2019erb, Escriva:2020tak, Escriva:2021aeh, Escriva:2022pnz, Escriva:2023uko},
\bea \label{avC}
  \!\!\!\!\!\!\! \avC(r) &\equiv& \frac{3}{R(r)^3} \int_0^{R(r)} \rmd \tilde R \tilde R^2 \C \\
  &=& - \frac{2}{r^3 e^{3\zeta(r)}} \int_0^r \rmd \tilde r \tilde r^2 e^{3\zeta} [ 2 \tilde r \zeta' + 3  ( \tilde r \zeta' )^2 + ( \tilde r \zeta' )^3 ] \ , \nonumber
\eea
where $R(t,r)=a(t) r e^{\zeta(t,r)}$, and $a(t)$ is the background scale factor. In the literature, the integral has been taken to the maximum of $\C(r)$; we keep the integration endpoint free and study the profile of $\avC(r)$.

The collapse threshold $\C_\thr$ or $\avC_\thr$ depends on the radial profile, varying from $0.4$ to $2/3$. It was observed in \cite{Harada:2015yda, Musco:2018rwt} that a smoother transition between the overdensity and the surrounding background corresponds to a lower threshold, while a sharper peak leads to a higher threshold, as pressure gradients resist collapse. In \cite{Atal:2019erb} it was argued that if $\C(r)$ has several peaks, the threshold should be compared to the value of $\C(r)$ at the first peak from the origin, as other peaks dissipate. In contrast, \cite{Nakama:2014fra} found that high-frequency modulation of a peak facilitates collapse, and \cite{Escriva:2023qnq} found that with two peaks, collapse can occur even if both are below the threshold.

\begin{figure*}
  \centering
  \includegraphics[scale=1]{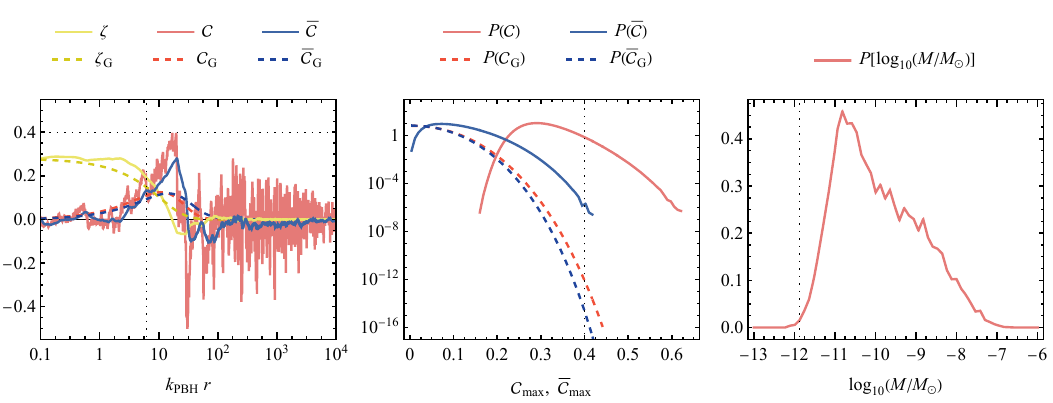}
  \caption{\emph{Left:} Solid lines show one realisation of the curvature perturbation $\zeta(r)$, the compaction function $\C(r)$ and its average $\avC(r)$. The realisation has been selected so that $\C_{\max}=0.4$. Dashed lines show the same quantities calculated from the mean profile assuming Gaussianity and the same maximum value of $\zeta(r)$. The horizontal line shows the collapse threshold 0.4. The vertical line marks the power spectrum peak scale. \\
  \emph{Middle:} Probability distribution of $\C_{\max}$ and $\avC_{\max}$. The solid lines show the results of our numerical computation, while the dashed lines correspond to the Gaussian mean case. The vertical line marks the collapse threshold 0.4. \\
  \emph{Right:} The mass distribution with the collapse threshold $\C_\thr=0.4$. The vertical line marks the power spectrum peak scale.}
  \label{fig:profiles_and_distributions}
\end{figure*}

In our case, gradients are large due to rapid stochastic fluctuations; a typical realisation of $\C(r)$ is shown in \fig{fig:profiles_and_distributions}. Because the compaction function is conserved on super-Hubble scales, the spikes persist until the fluctuation re-enters the Hubble radius, when the collapse starts. Then pressure gradients smooth out variations \cite{Germani:2023ojx}. Determining the damping and finding the collapse criterion for such sharp profiles requires a dedicated numerical analysis. We simply find the probability distribution for the maxima of $\C(r)$ and $\avC(r)$, called $\C_{\max}$ and $\avC_{\max}$ respectively, and discuss the PBH abundance for conventional values of the collapse threshold.

For comparison, we also find the abundance using the mean radial profile calculated assuming Gaussianity. The mean profile of a spherical fluctuation with maximum value $\zeta_0$ is $\av{\zeta(r)}=\frac{\zeta_0}{\sigma_\zeta^2} \int\frac{\rmd k}{k} \Pzeta(k) \frac{\sin(k r)}{k r}$, where $\sigma_\zeta^2\equiv\int\frac{\rmd k}{k} \Pzeta(k)$ \cite{Dekel:1981}. From $\av{\zeta(r)}$ we calculate $\C(r)$ and $\avC(r)$. The Gaussian distribution of $\zeta_0$ with variance $\sigma_\zeta^2$ then gives the distributions of $\C_{\max}$ and $\avC_{\max}$.

{\it From the inflaton potential to the compaction function.--}
To compute $\zeta$ stochastically, we split the inflaton field into long and short wavelength parts, $\phi = \bar\phi + \delta\phi$, separated by the coarse-graining scale $\kc\equiv\sigma a H_0$, where $H_0$ is the initial value of the Hubble parameter, and $\sigma \ll 1$ is a constant \cite{Starobinsky:1986fx, Morikawa:1989xz, Habib:1992ci, Mijic:1994vv, Starobinsky:1994bd}. The long-wavelength part $\phi$ is treated as a background, and evolves according to the following stochastic equations to leading order in the gradient expansion (see e.g. \cite{Pattison:2019hef, Figueroa:2021zah, Tomberg:2022mkt}):
\bea \label{bg_eom}
    \!\!\!\!\!\! \dot{\bar\phi} = \bar\pi + \xi_\phi \ , \ 
    \dot{\bar\pi} = -\qty(3 - \frac{1}{2}\bar\pi^2)\qty( \bar\pi + \frac{V_{,\bar\phi}(\bar\phi)}{V(\bar\phi)} ) + \xi_\pi \ ,
\eea
where $\dot{}\equiv\frac{\rmd}{\rmd N}$, and $\bar\pi$ is the mean field momentum, and we use units where the Planck mass is unity. The dynamics are determined by the potential $V$. We consider the Higgs-inspired inflation model of \cite{Figueroa:2020jkf, Figueroa:2021zah} that produces asteroid-mass PBHs while also fitting CMB observations \cite{Akrami:2018odb}. The perturbations are amplified when the inflation rolls over a local maximum in $V$. The effect of perturbations is described by the noise terms $\xi_\phi$ and $\xi_\pi$, caused by the short-wavelength modes crossing the coarse-graining scale. They are treated as linear perturbations, so the noise is Gaussian, with $\av{\xi_\phi(N)\xi_\phi(\tilde{N})} = \Pphi(N,\kc)\delta(N-\tilde{N})$, where $\Pphi(N,k) \equiv \frac{k^3}{2\pi^2}|\delta\phi_k(N)|^2$. Correlators that involve $\xi_\pi$ behave similarly.

In \cite{Figueroa:2020jkf, Figueroa:2021zah}, the coupled stochastic evolution of $\bar\phi$ and $\delta\phi$ was solved at every timestep. This is computationally expensive: it took over $10^6$ CPU hours to resolve the probability distribution up to $\zeta=0.95$. To be able to reconstruct the spatial profile of $\C(r)$, we simplify the calculation by using results from \cite{Tomberg:2023kli}.

Our PBH model, as is typical, first has a USR phase, when the inflaton climbs towards a local potential maximum, and then a dual constant-roll (CR) phase where it rolls down the other side \cite{Karam:2022nym}. The perturbations that exit the Hubble radius during USR get the largest amplitude. They exit the coarse-graining scale later, during the dual CR phase. We denote the wavenumber of the last mode that exits during USR by $\kpbh=2.0\times10^{13}$ Mpc$^{-1}$. During the dual CR phase the perturbations are adiabatic, meaning that the stochastic noise only moves the inflaton back and forth on the CR trajectory, where the perturbation evolution turns out to be independent of the noise \cite{Tomberg:2022mkt}. As long as this holds, the equations \eqref{bg_eom}  have the following analytical solution in terms of discrete $k$ modes \cite{Tomberg:2023kli}:
\begin{gather}
    \label{phi_solution}
    \phi(N) = \phi_0 e^{\frac{\epsilon_2}{2}N} \qty(1 - \frac{\epsilon_2}{2}X_{<\kc}) \, , \\
    \label{X_solution}
    X_{<k} \equiv -\!\sum_{\tk=\kini}^{k} \sqrt{\Pzeta(\tk) \, \rmd \ln k} \, \hxi_\tk \, , \quad \av{\hxi_k\hxi_\tk} = \delta_{k\tk} \, .
\end{gather}
Here $\phi_0$ is an integration constant, and $\epsilon_2 \equiv \frac{\rmd}{\rmd N} \ln (-\dot H/H)$ is the second slow-roll parameter during the dual CR phase (when it is constant). The power spectrum $\Pzeta(k)$ is pre-solved from the Mukhanov--Sasaki equation. In our model it peaks at $k_{\text{peak}}\approx0.2\kpbh$, and $\epsilon_2 = 0.807$ \cite{Figueroa:2021zah}, see \fig{fig:power_spectrum} in Appendix A.

The sum in \eqref{X_solution} is taken over $\tk$ values with fixed logarithmic step $\rmd \ln k$, which approaches zero in the continuum limit, starting from the initial coarse-graining scale $\kini=0.05$ Mpc$^{-1}= 2.5\times10^{-15} \kpbh$. The stochastic noises $\hxi_k$ are independent Gaussian random variables.

On the CR trajectory, the coarse-grained variable $\Delta N_{<k}=\zeta_{<k}$ is given in terms of $X_{<k}$ as \cite{Tomberg:2023kli}
\bea \label{zeta_vs_X}
    \zeta_{<k} = -\frac{2}{\epsilon_2}\ln(1 - \frac{\epsilon_2}{2}X_{<k}) \, .
\eea
The resulting probability distribution for $\zeta_{<k}$ matches the numerical result of \cite{Figueroa:2020jkf, Figueroa:2021zah} excellently, demonstrating the power of this technique.

The results \eqref{phi_solution}--\eqref{zeta_vs_X} characterise a single stochastic realisation completely in terms of the variables $\hxi_k$, a tremendous simplification compared to solving \eqref{bg_eom} by force. The model is fully described by $\Pzeta(k)$ and the parameter $\epsilon_2$ that controls non-Gaussianity. Combining \eqref{phi_solution}--\eqref{zeta_vs_X} with \eqref{zeta_k} and \eqref{zeta} and using It\^{o}'s lemma \cite{Ito:1944, Vennin:2015hra, Tomberg:2023kli} for differentiation of stochastic variables gives
\bea \label{r_zeta_prime}
    r\zeta'(r) &=& \sum_{\tk=\kini}^{\kend} \Bigg[ -\frac{\hxi_\tk}{1-\frac{\epsilon_2}{2}X_{<\tk}}\sqrt{\Pzeta(\tk) \, \rmd \ln k} \el
    && + \frac{\epsilon_2}{4\qty(1-\frac{\epsilon_2}{2}X_{<\tk})^2} \Pzeta(\tk) \, \rmd \ln k \Bigg] \el
    && \times\left[\cos{} ( \tk r ) - \frac{\sin{}(\tk r)}{\tk r}\right] \ .
\eea
The final mode with $\kend=2.6\times10^{16}$ Mpc$^{-1}= 1330\,\kpbh$ exits the Hubble radius about ten e-folds before the end of inflation. Plugging \eqref{r_zeta_prime} into \eqref{C} gives the profile $\C(r)$ in one patch. We collect statistics for the patches by generating many stochastic realisations $\hxi_k$.

The quantity $\C$ was first calculated from stochastic inflation in \cite{Tada:2021zzj}. The authors used the first passage time formalism to find the coarse-grained $\zeta_{<k}$, which they used to approximate $\zeta(r)$ coarse-grained over a region of size $\sim k^{-1}$. They computed the distribution of $\C_{\max}$ analytically in an example model, with the noise given in the slow-roll approximation. We obtain the radial profile in each patch, and compute the noise beyond the slow-roll approximation, which can make a large difference \cite{Figueroa:2020jkf, Figueroa:2021zah}.

{\it Results.--}
We generated $10^{9}$ stochastic realisations of $\zeta(r)$, each with 10450 $k$-points from $\kini$ to $\kend$, separated by step length $\rmd \ln k = 1/256$. We recorded the values of $\C_{\max}$ and $\avC_{\max}$, the values of $r$ for which they were reached, the value of $\Delta N_{<\kpbh} = \zeta_{<\kpbh}$ (denoted simply by $\Delta N$ below), and the number of peaks above $\C_\thr>0.4$ or $\avC_\thr>0.4$. We discuss resolution issues and the numerical algorithm in Appendices B and C.

The left panel in \fig{fig:profiles_and_distributions} shows a typical realisation of $\zeta(r)$, chosen so that $\C_{\max}=0.4$, and the corresponding $\C(r)$ and $\avC(r)$. The curvature perturbation is constant for small $r$, drops rapidly to zero around $r\sim10\,\kpbh^{-1}$, and oscillates stochastically before settling down. The oscillations are amplified in $\C(r)$, and its peak locations correspond to the big drop of $\zeta(r)$. The behaviour of $\avC(r)$ is similar, but smoother and more robust to small fluctuations. For example, it has only one high peak instead of two. The plots also show the mean profile $\av{\zeta(r)}$ in the Gaussian case with the same value of $\zeta(0)$, and the corresponding $\C(r)$ and $\avC(r)$.

In \cite{Escriva:2023qnq} a $\C(r)$ profile with two peaks was studied, with the caveat that such cases might be rare. We see that several peaks is the norm: 39\% of the simulations even have more than one peak above $\C_\thr=0.4$. The number $p$ of peaks above $\C_\thr=0.4$ follows an exponential distribution, $n(p)\propto e^{-\a p}$, with $\a=0.79$. See Appendix B for dependence on resolution. Perturbations in all patches are of type I \cite{Kopp:2010sh, Harada:2013epa}, \ie $R'>0$.

The middle panel in \fig{fig:profiles_and_distributions} shows the probability distribution of $\C_{\max}$ and $\avC_{\max}$. Averaging $\C$ suppresses the maximum values significantly. Also shown are the distributions calculated from the mean Gaussian profile. They fall off steeply, and do not reproduce the numerical distributions anywhere. Spherical symmetry, and thus our distributions, is not likely to be valid for small fluctuations, but they are not important for PBH formation.

The initial PBH abundance $\b$ is given by an integral over $P(\C_{\max})$ or $P(\avC_{\max})$ starting from the collapse threshold. For $\C_\thr=0.4$, we get $\b = 0.01$, while $\avC_\thr=0.4$ leads to $\b=5\times10^{-9}$. (The abundance for $\avC_{\max}$ has an error of some tens of \% because of the small number of realisations above $\avC_\thr$; in contrast, $\C$ is well sampled.) Using the Gaussian mean profile for the same thresholds instead gives $\b=4\times10^{-15}$ and $\b=9\times10^{-18}$, respectively. In \cite{Figueroa:2020jkf, Figueroa:2021zah}, we calculated $\beta$ for the same model from the $\Delta N$ distribution using the threshold $\zeta_\thr=1$, giving $\b=3\times10^{-11}$. (In the Gaussian approximation $\C_{\max}=0.4$ corresponds to $\zeta_0\approx1$, and $\avC_{\max}=0.4$ to $\zeta_0\approx1.1$.) The corresponding Gaussian linear perturbation theory result was $\beta=3\times10^{-16}$, tuned so that PBHs contribute all dark matter today.

\begin{figure}
\centering
\includegraphics[scale=1]{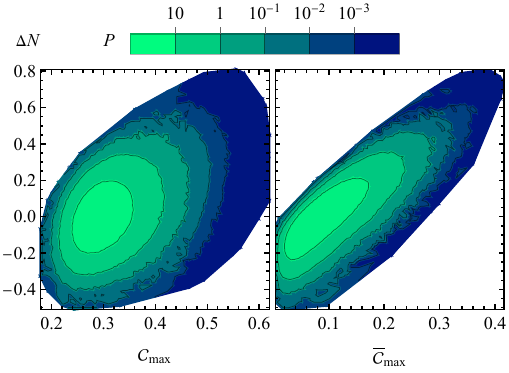}
\caption{Correlation of $\zeta=\Delta N$ with $\C_{\max}$ (left) and $\avC_{\max}$ (right).}
\label{fig:correlations}
\end{figure}

In \fig{fig:correlations} we show the correlation of $\Delta N$ with $\C_{\max}$ and $\avC_{\max}$. It demonstrates that $\Delta N$ is a poor predictor of $\C_{\max}$ and vice versa; many patches with $\C_{\max}>0.4$ even have $\Delta N<0$. The average $\avC_{\max}$ is better correlated with $\Delta N$, but even there the scatter is large, and we are limited by lack of statistics at large values of $\avC_{\max}$.

The PBH mass is roughly the mass inside the Hubble radius given by the peak scale $r$. There are corrections from evolution during the collapse and accretion \cite{Nakama:2013ica, Escriva:2019nsa, Gow:2020bzo, Escriva:2021pmf, Escriva:2023qnq}, including critical behaviour \cite{Niemeyer:1997mt, Niemeyer:1999ak, Musco:2008hv, Musco:2012au}. Neglecting these and assuming standard thermal history, the mass is (see \eg \cite{Karam:2022nym})
\bea \label{PBH_mass}
    M = 5.6 \times 10^{15} \qty(r \times 0.05 \text{ Mpc}^{-1})^{2} M_{\odot} \  .
\eea
The right panel of \fig{fig:profiles_and_distributions} gives the mass distribution normalised to unity using $\C_\thr=0.4$. It peaks around $M=1.5\times10^{-11} M_{\odot}$, compared to the monochromatic approximation in \cite{Figueroa:2020jkf, Figueroa:2021zah}, where the scale $\kpbh$ gave the mass $M=7\times10^{-15} M_{\odot}$. It has significant support over three orders of magnitude in mass, much broader than the distribution found in \cite{Tada:2021zzj}, stretching to higher values than that determined from the maximum of the power spectrum, whereas critical collapse extends the spectrum to smaller masses. For the threshold $\avC_\thr=0.4$, there are not enough realisations to obtain a reliable mass distribution, but the masses are of the same order.

{\it Conclusions.--}
We have for the first time constructed stochastic realisations of the spatial profiles of the curvature perturbation $\zeta$ and the compaction function $\C$, starting from the inflaton potential. The stochastic effects that have been shown to generate an exponential tail for the distribution of the curvature perturbation also lead to highly oscillatory profiles for the individual patches. Our calculation produces the individual profiles of each patch, allowing us to find the radial profiles of both $\C$ and the average compaction function $\avC$.

PBH abundance has sometimes been calculated assuming Gaussian statistics and using the maximum of $\zeta$ as the collapse criterion. Naively applying the PBH collapse threshold for $\C$ or $\avC$ that has been calibrated on smooth profiles to our spiky profiles enhances the PBH abundance by a factor of $10^9$ for $\C$ and a factor of $10^2$ for $\avC$, compared to using $\zeta$. Compared to calculating the compaction function from the mean profile, assuming the field to be Gaussian, as often done in the literature, the enhancement factor is $10^{13}$ for $\C$ and $10^9$ for $\avC$. 

However, the collapse criteria from simulations of smooth profiles cannot be reliably applied to the rapidly varying profiles we find, as shown by the fact that we have large differences between using $\C$ or $\avC$. On the one hand, having more than one spike can lead to collapse even if none of them exceed the threshold \cite{Escriva:2023qnq}. On the other hand, pressure gradients will dampen the spikes when the perturbation enters the Hubble radius and starts to collapse, which is expected to wash away some of the enhancement. Checking the collapse criterion anew with numerical simulations using the spiky profiles should establish which quantity correlates well with collapse, and the degree of profile-dependence of the collapse threshold. The enhancement factor can also be different for inflaton potentials corresponding to different PBH masses, as is the case for the exponential tail \cite{Figueroa:2021zah}.

The tail and spike enhancements could be important for determining whether the non-detection of PBHs rules out explaining the origin of gravitational waves possibly detected by pulsar timing arrays with second-order production from scalar perturbations \cite{Kalaja:2019uju, Gow:2020bzo}. With this mechanism, detection of the amplitude of gravitational waves would fix the amplitude of the scalar power spectrum. Because of the enhancements we have discussed, the power spectrum will lead to a larger PBH abundance, without changing the amplitude of gravitational waves. This can make the constraints tighter. It has also been argued that generation of PBHs in single-field inflation is excluded because the required large small-scale power spectrum would distort the power spectrum on CMB scales \cite{Inomata:2022yte, Kristiano:2022maq, Riotto:2023hoz, Choudhury:2023vuj, Kristiano:2023scm, Riotto:2023gpm, Firouzjahi:2023aum, Motohashi:2023syh, Firouzjahi:2023ahg, Franciolini:2023agm, Tasinato:2023ukp, Cheng:2023ikq, Fumagalli:2023hpa, Maity:2023qzw, Tada:2023rgp, Firouzjahi:2023bkt, Davies:2023hhn}. The effects we discuss would alleviate such tension, but it remains to be quantified how much smaller the power spectrum can be. We will report details of this and other extensions of our work elsewhere. \\

We thank Daniel Figueroa who participated in the beginning of the project. The authors wish to acknowledge CSC - IT Center for Science, Finland, for computational resources. This work was supported by the Estonian Research Council grant PRG1055 and by the EU through the European Regional Development Fund CoE
program TK133 ``The Dark Side of the Universe.'' E.T. was supported by the Lancaster--Manchester--Sheffield Consortium for Fundamental Physics under STFC grant: ST/T001038/1.

\bibliography{coll}

\clearpage

\section{Appendix A: curvature power spectrum}

\begin{figure}
  \centering
  \includegraphics[scale=1]{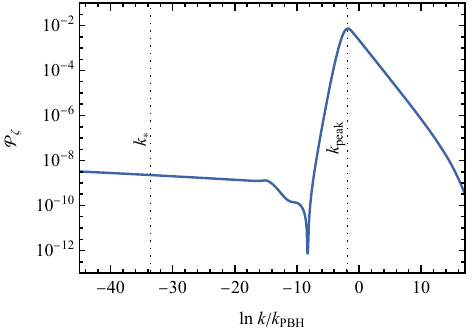}
  \caption{The curvature power spectrum. The vertical lines mark the peak scale $k_{\text{peak}}$ and the CMB pivot scale $k_*=0.05$ Mpc$^{-1}$.}
  \label{fig:power_spectrum}
\end{figure}

The inflationary model studied in this paper is the asteroid mass case previously studied in \cite{Figueroa:2020jkf, Figueroa:2021zah, Tomberg:2022mkt, Tomberg:2023kli}, built from Higgs inflation by adding quantum corrections to the plateau potential and tuning the CMB region by hand. The curvature power spectrum is depicted in \fig{fig:power_spectrum}, and it has a peak at $k= 0.2\kpbh = 3.3\times10^{12}$ Mpc$^{-1}$. After the peak, the power spectrum declines as $k^{-\epsilon_2}$, where $\epsilon_2 = 0.807$ is the second slow-roll parameter.

\section{Appendix B: details on resolution} 

We studied the effects of the $k$-resolution and the $r$-resolution used in our simulations on the PBH abundance, both separately and together.

To test the effect of the $k$-resolution, we considered four step lengths $\rmd \ln k$: $1/32$, $1/64$, $1/128$, and $1/256$. Reducing the step length appears to increase the number of peaks in $\C(r)$ at large values of $r$, shifting the $P(\C)$ peak to larger values of $\C$. The difference in the PBH abundance between the largest and the smallest step length is about a factor of 3. The relative difference decreases with each doubling of the $k$-resolution ($70$\%, $45$\% and $30$\%, respectively). This suggests that the abundance may eventually saturate with a small enough step length. The scaling of the number of peaks $n(p)\propto e^{-\a p}$ above the threshold $C_\thr=0.4$ varies with the resolution as  $\a=0.79$, $0.83$, $1.0$, and $1.4$, each value corresponding to halving our maximum resolution. Again, the changes decrease with increasing resolution. For the quantity $P(\avC)$, the effects of varying the $k$-resolution are negligible.

In our simulations, we considered $r$-values ranging from $10^{-1} \, \kpbh^{-1}$ to $10^{4} \, \kpbh^{-1}$. To test the effects of resolution in $r$, we considered three cases: $1024$, $2048$, and $4096$ $r$-points logarithmically distributed in the given range. We also considered extending the upper limit of the $r$-range by an order of magnitude to $10^{5} \, \kpbh^{-1}$. At the lowest $k$-resolution of $\rmd \ln k = 1/32$, the resolution or the range extension of $r$ have no quantifiable effect. However, higher $k$-resolution works in synergy with the increased $r$-resolution to enhance the peaks at large $r$, further pushing the $P(\C)$ peak towards larger values of $\C$. Doubling the $r$-resolution and extending the range of $r$ by an order of magnitude at the highest $k$-resolution $\rmd \ln k = 1/256$ increases the PBH abundance by $28$\%. As with the $k$-resolution, increasing the $r$-resolution and extending the range have negligible effect on $P(\avC)$.

The computational cost of the simulations is directly proportional to both the $k$- and $r$-resolution, so increasing both ramps up the amount of computational resources needed to produce sufficient statistics. In this Letter we have settled on using the highest tested $k$-resolution ($\rmd \ln k = 1/256$) with $1024$ $r$-points in the range spanning from $10^{-1} \, \kpbh^{-1}$ to $10^{4} \,  \kpbh^{-1}$.

\section{Appendix C: numerical algorithm}

Before the simulations begin, we pre-compute the power spectrum $\Pzeta(k)$ for our potential. Then, for each realisation, we execute the following numerical algorithm (simplified for readability):

\begin{algorithm}[H]
\SetAlgoVlined
\DontPrintSemicolon
\SetInd{0.5em}{1em}
\SetAlgoHangIndent{1em}
\SetVlineSkip{0.5em}
Set $X_{<0} = 0$.\;
\For{all modes k}
{
  Solve $\zeta_{<k}$ from \eqref{zeta_vs_X}.\;
  Generate Gaussian random variable $\hxi_k$.\;
  Solve $X_{<k + 1}$ from \eqref{X_solution}.
}
\For{all points r}
{
  Solve $\zeta(r)$ from \eqref{zeta}.\;
  Solve $r\zeta'(r)$ from \eqref{r_zeta_prime}.\;
  Solve $\mathcal C(r)$ from \eqref{C}.\;
  Solve $\avC(r)$ from \eqref{avC}.
}
Count the number of $\mathcal C(r) > 0.4$, $\avC(r) > 0.4$ peaks.\;
Store $\C_{\max}$, $\avC_{\max}$, the corresponding $r$ values, $\zeta_{<\kpbh}$, and the number of peaks.
\caption{Single realisation}
\end{algorithm} 

\end{document}